\documentclass[aps,pr,twocolumn,longbibliography,superscriptaddress]{revtex4-2}
\usepackage{bm}
\usepackage{color} 
\usepackage{graphicx}
\usepackage{epsfig}
\usepackage{natbib}
\usepackage{amsmath}
\usepackage{amssymb}
\usepackage{bbm}
\usepackage{multirow}
\usepackage{hhline}

\usepackage[T1]{fontenc}
\usepackage[utf8]{inputenc}
\usepackage{physics}

\usepackage[colorlinks=true,citecolor=blue,urlcolor=blue]{hyperref}

\newcommand{\Sec}[1]{{\color{blue}\it #1.} }

\newcommand{\rmi}{{\rm i}}

\newcommand {\e}{{\rm e}}

\begin{document}

\title{Coherent transport in non-Abelian quantum graphs}

\author{A.~V.~Poshakinskiy}
\email[E-mail:~]{poshakinskiy@gmail.com}
\affiliation{Ioffe Institute, St. Petersburg 194021, Russia}

\author{L.~E.~Golub}
\affiliation{Institute of Theoretical Physics and Halle-Berlin-Regensburg Cluster of Excellence CCE, University of Regensburg, 93040 Regensburg, Germany}

\begin{abstract}

We study quantum charge transport in two-dimensional networks in the presence of a magnetic field and spin-orbit interaction. The interplay of the corresponding Abelian and non-Abelian gauge fields leads to an intricate behavior of the conductance, which has different periodicities in the diffusive and ballistic regimes. We classify all configurations of magnetic and spin-orbit fields where a logarithmically divergent weak-(anti)localization correction appears in the diffusive regime. The conductivity of topologically distinct configurations is the same in the diffusive regime but different in the ballistic regime. The proposed setup provides a feasible realization of quantum graphs with non-Abelian gauge fields.

\end{abstract}

\maketitle

\Sec{Introduction}%
Coherent transport phenomena provide a powerful framework for realizing quantum effects in conducting systems.
In ring-like devices, the phase difference between the two interfering electron paths can be controlled by a transverse magnetic field via the Aharonov--Bohm (AB) effect~\cite{AltshulerAronov1985,Chakravarty1986}, or by a transverse electric field that induces spin-orbit interaction via the Aharonov--Casher (AC) effect~\cite{Aronov1993,Nitta1999,Meijer2002,Molnar2004,Frustaglia2004,Capozza2005,Hijano2021}. This leads to conductivity oscillations as a function of the magnetic field or spin-orbit interaction strength.

AB and AC oscillations are suppressed by electron scattering that destroys the interference for most of the electron trajectories. Nevertheless, in diffusive samples, there remains interference between the loop paths that an electron can follow in the two opposite directions around the ring. It leads to a weak-localization correction to the ring conductivity that oscillates with the magnetic field with half the period expected from the AB effect~\cite{AltshulerAronov1985,Chakravarty1986,Aronov1987}. To enhance the conductivity oscillations, networks with many rings were proposed~\cite{Doucot1985,Texier2004,Texier2009} and realized~\cite{Pannetier1984,Pannetier1985,Naud2001,Ferrier2004,Sawada2018}.
A similar change of the AC-oscillation period in the diffusive regime might be expected, but the AC oscillations in the diffusive regime have not been explored so far, with the existing studies limited to the case of weak spin-orbit interaction in rings and tubes~\cite{Engel2000,Kammermeier2016}.

In this work, we consider diffusive semiconductor networks and reveal conductivity oscillations as a function
of spin-orbit interaction strength. We demonstrate their intricate relationship with AC oscillations in the ballistic regime, due to the non-Abelian nature of the spin-orbit interaction. The proposed structures provide a feasible realization of {\it quantum graphs}~\cite{bookBerkolaiko2012}, i.e., metric graphs whose edges are equipped with a self-adjoint differential operator, where the spin-orbit interaction gives rise to a $\operatorname{SO}(3)$ non-Abelian gauge field~\cite{Frohlich1993,Hatano2007,Chen2008,Yang2008,Tokatly2008,Tokatly2010a} on the graph. 
We describe all parameter classes that lead to logarithmic divergences in the weak-localization correction to the network conductivity. They include, in particular, the class where the gauge field on the graph reduces to a pure gauge. Because the SO(3) group is not simply connected, such pure gauges can be of two distinct topological sub-classes~\cite{Poshakinskiy2025}, which yield the same conductivity in the diffusive regime, but a different ballistic conductivities, determined by the lift of the gauge field to the $\operatorname{SU}(2)$ group. 


\begin{figure}[b!]
    \centering
    \includegraphics[width=0.8\columnwidth]{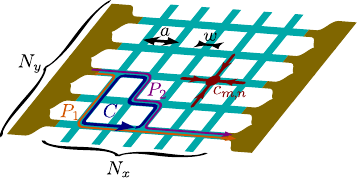}
    \caption{Schematic of a square network connected to the left and right leads. The incoming flux for a network vertex is shown in red. Loop contour $C$ shows the interference contribution for a pair of ballistic paths $P_1$ and $P_2$. 
}
    \label{fig:0}
\end{figure}

\begin{figure*}[t!]
    \centering
    \includegraphics[width=0.8\textwidth]{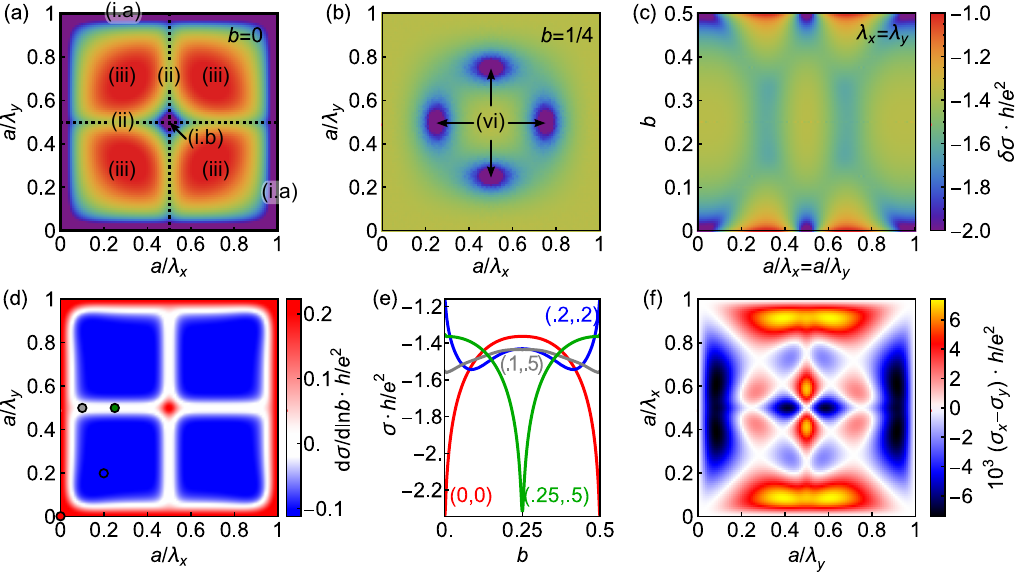}
    \caption{(a,b,c) Quantum correction to the conductivity of the diffusive network as a function of spin-orbit parameters and magnetic field. The configurations (i)--(iv) that support long-lived cooperon modes are labeled. (d) The magnetic-field derivative of the conductivity $d\sigma/d\ln b$ calculated at $b=0.01$ as a function of spin-orbit parameters. (e) The magnetic-field dependence of the conductivity for four different configurations of spin-orbit parameters $(a/\lambda_x,a/\lambda_y)$, indicated in the graph and shown by the colored dots in panel (d), which belong to classes (i)--(iv). (f) The conductivity anisotropy as a function of the spin-orbit parameters at zero magnetic field. All calculations are performed for $\kappa = 0.1$.
}
    \label{fig:WL}
\end{figure*}
 
\Sec{Model}
We consider a two-dimensional square network with period $a$ formed by narrow channels of width $w \ll a$, see Fig.~\ref{fig:0}. Within the channels, the degenerate electron gas is described by the Hamiltonian
\begin{align}\label{eq:H}
    H = -\frac{\hbar^2}{2m^*} \sum_{\alpha=x,y}\left(\frac{\partial}{\partial r_\alpha}- \rmi {A}_\alpha \right)^2  ,
\end{align}
where the gauge field $\bm A$ comprises the effects of $\bm k$-linear spin-orbit interaction and magnetic field:
${A}_\alpha  = \tfrac12  \bm\Lambda_{\alpha} \cdot \bm\sigma + e B x\, \delta_{\alpha,y}$,
$\bm\sigma$ is the vector of Pauli matrices, $m^*$ is the electron effective mass and $B$ is the strength of the magnetic field directed perpendicular to the network. Vectors $\bm\Lambda_{x(y)} = 2\pi \bm e_{y(x)}/\lambda_{x(y)}$ describe the spin-orbit coupling, where the spin precession length $\lambda_{x(y)}$ is expressed via Rashba ($\alpha$) and Dresselhaus ($\beta$) parameters as $\lambda_{x(y)} = \pi \hbar^2/[m^*(\beta \pm \alpha)]$ for (001)-oriented structures with $x \parallel [1\bar{1}0]$, $y \parallel [110]$, and $z \parallel [001]$~\cite{Ganichev2014}.
We consider weak magnetic fields where the Zeeman effect can be neglected.

We start with the diffusive regime, where the electrons exhibit multiple collisions when propagating along a network edge, $w,a \gg v_F \tau$, where $v_F$ is the electron Fermi velocity and $\tau$ is the scattering time. Then, the electron transport within the edges is described by a 2D diffusion coefficient $D= v_F^2\tau/2$ and the dominant contribution to the conductivity of a single edge is given by the Drude formula $\sigma_{\rm Dr} = n e^2 \tau/m^*$, where $n$ is the electron density inside the edge. The conductivity of a network with unit area then reads $\sigma_{0} = \sigma_{\rm Dr} w/a$. This classical result 
is independent of both spin-orbit interaction and magnetic field (in the regime $eB\tau/m^*c \ll 1$). However, such dependence arises in a quantum correction caused by the interference of electrons propagating along loop paths in opposite directions.

\begin{table*}
\caption{Classification of the configurations that have divergent conductivity corrections in the diffusive regime. Stars denote generic values. Long-lived modes are indicated by the value 1 of the corresponding Wilson loop operator $\mathcal{U}^{(\nu)}(C)$ or $U(C)$ for any number of plaquettes $P_C$ enclosed by the contour $C$. Conductivity correction in the diffusive regime $\delta\bar\sigma$ is expressed in terms of $\sigma_0=(e^2/\pi h)K(1/\cosh \kappa)\tanh(\kappa)/\kappa \sim (e^2/\pi h)\ln (l_\phi/a)$ at $l_\phi \gg a$. Column $\Delta\sigma(B)$ shows the sign of the conductivity change when the magnetic field is detuned.}
\label{tbl:cls}
\begin{ruledtabular}
\begin{tabular}{lcc cccccc cr}
\multirow{2}{*}{Class} & $(a/\lambda_x,a/\lambda_y)$ & $Ba^2/\Phi_0$
& \multicolumn{6}{c}{Diffusive transport}
& \multicolumn{2}{c}{Ballistic transport} \\
\cline{4-9}\cline{10-11}
& mod 1 & mod 1
& $\mathcal{U}^{(s)}(C)$
& $\mathcal{U}^{(t)}_{xx}(C)$
& $\mathcal{U}^{(t)}_{yy}(C)$
& $\mathcal{U}^{(t)}_{zz}(C)$
& $\delta\bar\sigma$
& $\Delta\sigma(B)$
& $U(C)$
& $\sigma$ \\[.3ex]
\colrule
i.a    & \multirow{2}{*}{$(0,0)$} & 0
       & \multirow{4}{*}[-1ex]{$1$}
       & \multirow{4}{*}[-1ex]{$1$}
       & \multirow{4}{*}[-1ex]{$1$}
       & \multirow{4}{*}[-1ex]{$1$}
       & \multirow{4}{*}[-1ex]{$-2\delta\sigma_0$}
       & \multirow{4}{*}[-1ex]{$>0$}
       & 1 & max \\
i.a$'$ & & $\frac12$
       & & & & & & & $(-1)^{P_C}$ & low \\
i.b    & \multirow{2}{*}{$(\frac12,\frac12)$} & 0
       & & & & & & & $(-1)^{P_C}$ & low \\
i.b$'$ & & $\frac12$
       & & & & & & & 1 & max \\[.3ex]
\colrule
ii  & $(\frac12,\ast)$ & $0$ or $\frac12$ & 1 & 1 & $\ast$ &  $\ast$ & $\sim \text{const}$   & $\approx 0$ & $\ast$  & $\ast$  \\
iii & $(\ast,\ast)$   & $0$ or $\frac12$ & 1 &  $\ast$ &  $\ast$ & $\ast$  & $\sim \delta\sigma_0$ & $<0$ &  $\ast$ & $\ast$  \\
iv  & $(\frac12,\frac14\text{ or }\frac34)$ & $\frac14$ or $\frac34$ & $(-1)^{P_C}$ & $(-1)^{P_C}$ & 1 & 1 & $-2\delta\sigma_0$ & $>0$ & $\ast$  &  $\ast$ \\
\end{tabular}
\end{ruledtabular}
\end{table*}

The quantum correction to the local conductivity is given by the difference of the triplet and singlet components of the cooperon $\bm C^{(t)}$ and $C^{(s)}$: $\Delta\sigma(\bm r) =  (2De^2/h) [C^{(s)}(\bm r,\bm r)-{\rm Tr\,} \bm C^{(t)}(\bm r,\bm r) ]$~\cite{AltshulerAronov1985,Chakravarty1986,Hikami1980,Iordanskii1994,Glazov2009,Saito2022,Golub2024}. The cooperon is found from the diffusion-like equations
\begin{align}\label{eq:diff}
D\left[\frac{1}{l_{\phi}^2} -  \sum_{\alpha  = x,y} \left(\frac{\partial}{\partial r_\alpha} - \rmi {\mathcal A}^{(\nu)}_\alpha \right)^2  \right] C^{(\nu)}(\bm r,\bm r') = \delta(\bm r- \bm r'),
\end{align}
which feature the $\operatorname{U}(1)$ gauge field ${\mathcal A}^{(s)}_{\alpha} = 2 e B x\, \delta_{\alpha,y}$ for the singlet component, and the $\operatorname{U}(1)\times \operatorname{SO}(3)$ gauge field ${{\mathcal{A}}}^{(t)}_{\alpha} \bm C^{(t)}=  2 e B x\, \delta_{\alpha,y}\, \bm C^{(t)} - \rmi  \bm\Lambda_{\alpha}\times \bm C^{(t)}$ for the triplet component. We also introduced here the phase relaxation length $l_\phi$.
Equation~\eqref{eq:diff} must be supplemented with the Neumann boundary conditions that correspond to the absence of charge and spin fluxes through the boundary: $\bm n \cdot \bm J^{(\nu)} = 0$, where  $\bm n$ is the normal vector to the boundary and $J_\alpha^{(\nu)} = ({\partial}/{\partial r_\alpha} - \rmi{\mathcal A}^{(\nu)}_\alpha)  C^{(\nu)}$. Then, the left-hand side of Eq.~\eqref{eq:diff} is a self-adjoint operator with real positive eigenvalues that we denote as $\gamma^{(\nu)} = D{k^{(\nu)}}^2$. 

To solve Eq.~\eqref{eq:diff} in the network, we exploit the separation of length scales $v_F\tau \ll w \ll a, \lambda_{x(y)},l_\phi$. Then, for $\bm r$, $\bm r'$ inside a horizontal network edge and $|\bm r- \bm r'|  \ll a$, the cooperon reads
$ C^{(\nu)}(\bm r, \bm r')  = C^{(\nu)}_{\rm ch}(x-x';y,y') + C^{(\nu)}_{\rm net}(x,x')$, where
$C^{(\nu)}_{\rm ch}$ is the solution of the diffusion equation~\eqref{eq:diff} in the infinite horizontal channel, and the contribution $C^{(\nu)}_{\rm net}$ is due to the effect of the vertices and the other edges of the network. The quantum corrections to the conductivity of infinite channels were analyzed in detail in Ref.~\cite{Wenk2010}. Importantly, the triplet part of the cooperon in the channel is affected by the Dyakonov-Perel spin relaxation, which, for a channel of width $w \ll \lambda_{x,y}$, is described by the relaxation-rate tensor $\bm\Gamma_{\rm ch}=D w^2(\Lambda_\perp^2 - \bm\Lambda_\perp \otimes \bm\Lambda_\perp)/12$~\cite{Malshukov2000,Kiselev2000,Wenk2011,Altmann2015,Grobecker2025}, where $\bm\Lambda_\perp = \bm\Lambda_x \times \bm\Lambda_y$, and leads to a transition to weak antilocalization with increasing $w$~\cite{Kettemann2007,Wenk2010}. Here, for simplicity, we consider the case of sufficiently narrow channels, so $\Gamma_{\rm ch}  \ll D/l_\phi^2$ and the Dyakonov-Perel spin relaxation inside the edges can be neglected. Then, $C^{(\nu)}_{\rm ch}$ yields a negative contribution to the conductivity that is independent of spin-orbit parameters and the magnetic field provided $B \ll \Phi_0/(wl_\phi)$; we disregard it in what follows. 
Instead, we focus on the contribution $C^{(\nu)}_{\rm net}$, which can be found by solving the 1D counterparts of Eq.~\eqref{eq:diff}, obtained by keeping only $\alpha = x(y)$ in each horizontal (vertical) edge and by matching the solutions at the network vertices. Mathematically, such a structure defines a quantum graph with Kirchhoff boundary conditions~\cite{bookBerkolaiko2012}.

The eigenfunctions $c^{(\nu)}$ of these equations form two families.
Those from the first family take nonzero values at the vertices. 
Then, following the de Gennes--Alexander approach~\cite{Alexander1983,Rammal1983,Doucot1986}, we express the flux coming to the vertex $(m,n)$ from the adjacent vertex $(m+1,n)$ as,
\begin{align}
J^{(\nu)} = \frac{D k^{(\nu)}}{\sin k^{(\nu)}a}\big(  \mathcal{U}^{(\nu)\dag}_{m,n;x}  c^{(\nu)}_{m+1,n} - c^{(\nu)}_{m,n} \cos k^{(\nu)}a \big) ,
\end{align}
where $\mathcal{U}_{m,n;x}^{(\nu)} = \e^{\int_0^a \rmi\mathcal A_x (ma+x,na) dx} $ is the Wilson link operator,
and similarly for the other adjacent vertices. The total flux into to the vertex (see the red arrows in Fig.~\ref{fig:0}) must vanish, leading to the eigenvalue problem
$\sum_{m',n'} H_{m,n;m',n'}^{(\nu)}c^{(\nu)}_{m',n'} = 4 \epsilon^{(\nu)} c^{(\nu)}_{m,n}$, where 
\begin{align}\label{eq:tight}
H_{m,n;m',n'}^{(\nu)} = 
&\mathcal{U}_{x;m-1,n}^{(\nu)} \delta_{m',m-1}\delta_{n,n'}+ \mathcal{U}_{x;m,n}^{(\nu)\dag} \delta_{m',m+1}\delta_{n,n'} \nonumber \\
+ \mathcal{U}_{y;m,n-1}^{(\nu)}&\delta_{m,m'}\delta_{n',n-1} + \mathcal{U}_{y;m,n}^{(\nu)\dag} \delta_{m,m'}\delta_{n',n+1}\,,
\end{align}
and $\epsilon^{(\nu)} = \cos k^{(\nu)} a$ plays the role of the eigenenergy. Hamiltonian~\eqref{eq:tight} describes the standard tight-binding problem for a particle subjected to the gauge field.
There exists also the second family of the eigenfunctions that are zero at all vertices, and can be chosen to be localized on the cycles (unit cells) of the network~\cite{bookBerkolaiko2012}. The corresponding eigenvalues have flat dispersion $k^{(\nu)}=\pi p/a$, with positive integer $p$, and are independent of the magnetic field and spin-orbit interaction.

First, we analyze the isotropic part of the network conductivity correction, $\delta\bar\sigma = (\delta\sigma_{xx} + \delta\sigma_{yy})/2$, and later discuss the smaller anisotropic correction $\sigma_a = \delta\sigma_{xx} - \delta\sigma_{yy}$.
The isotropic correction $\delta\bar\sigma$ is determined by the cooperon value averaged over the coordinates, $\bar C^{(\nu)} = \sum 1/(2wa N_c D k^{(\nu)2})$, where $N_c$ is the number of unit cells in the network and the summation is taken over all the eigen wave vectors $k^{(\nu)}$. The contribution of the dispersive modes is simplified by expressing $k^{(\nu)} = (\arccos \epsilon^{(\nu)} + 2\pi p)/a$ and performing the summation over integer $p$, which gives
\begin{align}
\delta\bar\sigma = \frac{e^2 }{h \kappa N_c} \sum_{\epsilon=\epsilon^{(s)},\epsilon^{(t)}} \frac{\pm \sinh \kappa}{\cosh \kappa - \epsilon}\,,
\end{align}
where $+$($-$) is taken for the eigenvalues $\epsilon^{s(t)}$ and $\kappa = a/l_\phi$. 
%
The flat modes give the conductivity contribution $\delta\bar\sigma_0 = - (e^2/h)(\kappa \coth \kappa -1)/\kappa^2$,  which is independent of the magnetic field and spin-orbit interaction, and is finite at $\kappa \to 0$; we disregard it in what follows.

\Sec{Classification of spin-orbit coupled networks}
Figure~\ref{fig:WL} presents the network conductivity correction as a function of the spin-orbit interaction strength, characterized by the number of full turns the electron spin makes while traveling along the horizontal (vertical) edge $a/\lambda_{x(y)}$, and of the magnetic-field strength, characterized by the reduced AB phase that an electron acquires when traveling around the plaquette  $b=Ba^2/\Phi_0$, where $\Phi_0 = hc/e$ (for the cooperon, the reduced phase is $2b$). The plotted range corresponds to a single period of the AB and AC oscillations: a period of $1/2$ in $b$ and a period of 1 in $a/\lambda_{x(y)}$. 
Note that in realistic networks with finite edge width $w$, the period of the AB and AC oscillations is smeared and they decay at $b \gtrsim a/w$~\cite{Doucot1986} and $a^2/(\lambda_x\lambda_y) \gtrsim a/w$, respectively.

\begin{figure*}[t!]
    \centering
    \includegraphics[width=0.8\textwidth]{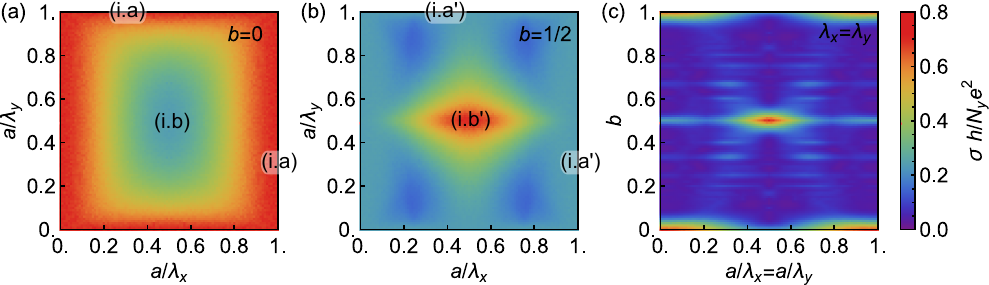}
    \caption{Conductivity of the ballistic network as a function of spin-orbit parameters for (a) $b=0$, (b) $b=1/2$, and (c) as a function of magnetic field and spin-orbit strength for $a/\lambda_x=a/\lambda_y$. The calculation is performed for $N_y = 10$, $N_x \to \infty$. 
    }
    \label{fig:Land}
\end{figure*}

The effect of the gauge fields on the cooperon can be understood through the $\operatorname{U}(1) \times \operatorname{SO}(3)$ Wilson-loop operators 
\begin{align}\label{eq:loop}
   \mathcal{U}^{(\nu)}(C) =  \mathcal{P}\exp\oint\limits_C \sum_{\alpha=x,y}\rmi\mathcal{A}^{(\nu)}_\alpha dr_\alpha
\end{align}
where the integral is taken along a loop $C$ that encloses one or several plaquettes (see the blue path in Fig.~\ref{fig:0} as an example) and $\mathcal P$ denotes path ordering. If there is a state that remains unchanged under $\mathcal{U}^{s(t)}(C)$ for any $C$, a long-lived cooperon mode emerges and gives rise to a conductivity contribution  $\pm \delta\sigma_0=\pm (e^2/\pi h)K(1/\cosh \kappa)\tanh(\kappa)/\kappa$~\cite{Doucot1986} ($K$ is the complete elliptic integral of the first kind) that diverges logarithmically as $ \delta\sigma_0 \sim -(e^2/\pi h)\ln \kappa$ at $\kappa \to 0$.  
All configurations that support long-lived cooperon modes can be organized into the classes, listed in Table~\ref{tbl:cls}.

Classes (i)--(iii) are realized at $b=0$ (mod $1/2$). In class (i), the gauge field can be eliminated by a gauge transformation. All cooperon modes are long-lived, indicated by the values 1 of the $\mathcal{U}^{(\nu)}(C)$ components, resulting in a net negative logarithmic conductivity correction, see the dark purple color in Fig.~\ref{fig:WL}(a). In class (ii), contributions to the conductivity from the long-lived singlet mode and the only long-lived triplet mode cancel each other, yielding no logarithmic divergence. A smaller negative contribution to the conductivity comes from the short-lived triplet modes, see the green color in Fig.~\ref{fig:WL}(a).  In class (iii), only the singlet long-lived mode survives, yielding a positive logarithmic correction to the conductivity, see the red color in Fig.~\ref{fig:WL}(a).
Class (iv) occurs at $b=1/4$ (mod $1/2$) and features two long-lived triplet modes. Interestingly, the contributions of the remaining short-lived triplet mode and the singlet mode cancel each other precisely. The resulting conductivity correction is the same as in class (i), see the dark purple color in Fig.~\ref{fig:WL}(b).

In class (i), topological properties can be associated with the triplet Wilson loops~\cite{Poshakinskiy2025}. Consider the Wilson line operator $\mathcal{U}^{(t)}(C,l)$, defined similarly to Eq.~\eqref{eq:loop} but with the integral taken along the initial part of $C$ with length $l \in [0,L_C]$, where $L_C$ is the length of $C$. Since $\mathcal{U}^{(t)}(C,0)$ and $\mathcal{U}^{(t)}(C,L_C)$ both equal identity, the function $\mathcal{U}^{(t)}(C,l)$ defines a loop path in the $\operatorname{SO}(3)$ group. Such paths can be classified according to the fundamental group of $\operatorname{SO}(3)$, which is $\mathbb{Z}_2=\{+1,-1\}$. It turns out that the classes (i.a) and (i.b) are topologically distinct: for (i.a) all Wilson loops belong to the $+1$ class, while for (i.b) the class of each Wilson loop is determined by $(-1)^{P_C}$.
While in the diffusive regime the conductivity is identical in the classes (i.a) and (i.b), we shall see that the difference arises in the ballistic regime.

Now we turn to the magnetic-field dependence of the conductivity. The long-lived modes of the cooperon are destroyed by a slight detuning of the magnetic field. This leads to a sharp suppression of the conductivity correction by the magnetic field, see Fig.~\ref{fig:WL}(c). Figure~\ref{fig:WL}(e) shows the magnetic-field dependencies for representative configurations from classes (i)--(iv). A weak-localization-like behavior occurs around $b=0$ in class (i) (red curve) and around $b=1/4$ in class (iv) (green curve). A typical non-monotonic weak-antilocalization-like behavior occurs around $b=0$ in class (iii) (blue curve). In class (ii) (gray curve), the magnetic field dependence is weak. This suggests a method to discriminate between the different cases through the sign of the conductivity change under a small magnetic field variation, as is demonstrated in Fig.~\ref{fig:WL}(d) for $b$ around 0. The positive conductivity variation (red color) indicates class (i), the negative one (blue color) indicates class (iii), while no variation (white color) indicates class (ii).

Finally, we discuss the network conductivity anisotropy $\sigma_a = \sigma_{xx} - \sigma_{yy}$. Due to anisotropic spin-orbit interaction, the triplet cooperons $\bm C^{(t)}(\bm r,\bm r)$ for $\bm r$ in the horizontal and vertical edges of the network can differ, giving rise to a finite $\sigma_a$, as calculated in Appendix~\ref{App:A}. In contrast to the isotropic part of the conductivity, $\sigma_a$ is always finite at $\kappa \to 0$. The dependence of $\sigma_a$ on the spin-orbit strength for $b=0$ is shown in Fig.~\ref{fig:WL}(f). As expected, $\sigma_a$ vanishes if $\lambda_x=\lambda_y$, when the spin-orbit interaction is isotropic, and also in class (i), when the spin-orbit interaction can be gauged away.

\Sec{Comparison with ballistic networks}
If the electrons exhibit no scattering while propagating through the network, their wave functions can be obtained by solving the 1D Schr\"odinger equations with the Hamiltonian of Eq.~\eqref{eq:H}, where only $\alpha=x(y)$ is kept for horizontal (vertical) edges. At the vertices, for the sake of simplicity, we impose the Kirchhoff boundary conditions~\cite{Griffith1953,Xia1992}: the continuity of the wave function $\psi(\bm r)$ and zero total flux of $(\partial/\partial\bm r - \rmi\bm A)\psi$ into the vertex.
This yields a system of equations for the spinor wave function at the lattice vertices $\psi_{m,n}$ of exactly the same form as Eq.~\eqref{eq:tight}, but with the  $\operatorname{U}(1) \times \operatorname{SU}(2)$ gauge field defined through $\bm A$ instead of $\bm {\mathcal A}$~\cite{Vidal2000,Bercioux2005}, and $k=\sqrt{2m^*E}/\hbar$, where $E$ is the electron energy. Assuming that the temperature $T \gtrsim \hbar^2/(m^*a^2)$, we average over $k$. 

To calculate the conductance of the $N_x\times N_y$ ballistic network, we use the Landauer--B\"uttiker formula,
\begin{align}\label{eq:sigB}
    \sigma = \frac{e^2}{hN_y} \sum_{n, n'=1}^{N_y} \operatorname{Tr}(s_{n',n}^\dag s_{n',n} )\,,
\end{align}
where the $2\times 2$ matrices $s_{n',n}$  
are computed as sums of transmission amplitudes for all possible paths connecting leads $n$ and $n'$.
To calculate numerically the conductance dependence on spin-orbit parameters and magnetic field strength, shown in Fig.~\ref{fig:Land}, we use the transfer-matrix technique outlined in Appendix~\ref{App:B}.

After averaging Eq.~\eqref{eq:sigB} over $k$, there remains only the interference between paths of the same length (see $P_1$ and $P_2$ in Fig.~\ref{fig:0}), which can be regarded as two halves of a loop path $C = P_2^{-1}P_1$. 
The conductivity is then determined by the sum of $\operatorname{Re} \operatorname{Tr} U(C)$ over all the loops, with certain coefficients, where the Wilson loop operators $U(C)$ are defined similarly to Eq.~\eqref{eq:loop}, but with the $\operatorname{U}(1) \times \operatorname{SU}(2)$ gauge field $\bm A$.
Now compare this with the diffusive result, which can also be rewritten as a sum over loops of $\operatorname{Re} \operatorname{Tr}U^2(C)$~\cite{Chakravarty1986}.
In case of a pure magnetic field, the gauge field is Abelian, and $U^2(C)$ corresponds to just a twice larger magnitude of the field than $U(C)$, which leads to a twice shorter period of AB oscillations in the diffusive regime as compared to the ballistic regime~\cite{AltshulerAronov1985,Chakravarty1986}. However, this is not the case for the non-Abelian spin-orbit gauge field, so the period of the AC oscillations is the same in the ballistic and diffusive regimes. 

The $\operatorname{U}(1) \times \operatorname{SU}(2)$ loop operators $U(C)$ can be regarded as the lift of the previously classified $\operatorname{U}(1) \times \operatorname{SO}(3)$ operators $\mathcal U^{(t)}(C)$. Importantly, the lift of the same $\mathcal U^{(t)}(C)$ can yield distinct values of $U(C)$ that differ by a sign. In particular, this occurs for the topologically distinct  classes (i.a) and (i.b), which both have $\mathcal U^{(t)}(C)=1$, but the lift yields $U(C) = 1$ in class (i.a) and $U(C) = (-1)^{P_C}$ in class (i.b), see Table~\ref{tbl:cls}. As a result, the ballistic conductance is distinct in the two cases, in contrast to the weak-localization correction, cf. Figs.~\ref{fig:Land}(a) and~\ref{fig:WL}(a). 
The highest ballistic conductance is achieved in class (i.a), where $b=0$ and the spin-orbit interaction can be gauged away, and in class (i.b$'$), where the spin-orbit interaction gives rise to an additional AC phase $\pi$ that cancels the AB $\pi$ phase for any plaquette. In classes (i.b) and (i.a$'$), either an AB or AC phase of $\pi$ leads to  destructive interference of electron paths and suppressed conductance.

The ballistic conductance always decreases with $b$ at small $b$, see Fig.~\ref{fig:Land}(c), in contrast to the conductivity in the diffusive regime. Moreover, the dependence on $b$ is not smooth (in the limit $N_{x},N_y \to \infty$), as it is determined by the fractal energy band structure of the lattice Hamiltonian~\eqref{eq:tight}, similar to the Hofstadter butterfly~\cite{Hofstadter1976}. 

\Sec{Outlook}
The considered spin-orbit coupled networks provide a feasible realization of quantum graphs with non-Abelian gauge fields. Such structures can be fabricated from semiconductor quantum wells using optical or electron-beam lithography~\cite{Altmann2015,Eberle2021}. The spin-orbit interaction, determining the gauge field, can be fully controlled via the voltages at the top and back gates~\cite{Dettwiler2017,Chander2026}. 

Realistic structures operate in the diffusive regime. High-mobility structures are expected to lie at the crossover to the ballistic regime~\cite{Ishihara2022} and might demonstrate a combination of the features predicted for the two regimes: the doubling of the AB-oscillation period~\cite{Bardarson2010} and the change in the AC-oscillation shape predicted here. Fully ballistic propagation can be realized in a network of $p$-wave superconductors with spin-orbit interaction, although the spin-orbit field there is described by the $\operatorname{SO}(3)$ group, making such systems similar to the diffusive networks considered here. We predict that the conductivity of such a superconducting network at near-critical temperatures will oscillate with the strength of the spin-orbit interaction, similarly to the 
Little--Parks conductivity oscillations with magnetic field~\cite{Sochnikov2010,Wang2025}.

\Sec{Acknowledgments}%
%
%
The work of L.E.G. was funded by the German Research Foundation (DFG) as part of the German Excellence Strategy -- EXC3112/1 -- 533767171 (Center for Chiral Electronics).

\appendix

\section{Anisotropy of the weak-localization correction to network conductivity}
\label{App:A}

When a voltage along the $x$ direction is applied to the network, only the horizontal edges contribute to the conductivity. To determine the network conductivity correction in such a configuration, we need to average $\delta\sigma_x = (1/a)\int_0^a \delta\sigma_x(x) \,dx$, where $ \delta\sigma_x(x) = (w/a)\Delta\sigma(x,0)$, and 
$\Delta\sigma(\bm r) = \frac{2e^2D}{ha} [C^{(s)}(\bm r,\bm r)-{\rm Tr\,}\bm C^{(t)}(\bm r,\bm r)] $ is the local conductivity correction. The cooperon is expressed via the eigenmodes as
$ C^{(\nu)}(\bm r,\bm r) = \sum_{k^{(\nu)}} |c^{(\nu)}(\bm r)|^2 /(N_c D k^{(\nu)2}) $.

We start by calculating the contribution of dispersive modes, which take nonzero values in the vertices. The eigenfunction value inside an edge can be expressed via its values at the vertices connected by the edge:
\begin{align}\label{eq:cx}
   &c^{(\nu)}(x) = \frac{1}{\sqrt{2wa}}\Big[ 
   \e^{\int_0^x \rmi\mathcal{A}_x^{(\nu)}(x') dx'} c^{(\nu)}(0)\, \sin k^{(\nu)}(a-x) 
   \nonumber \\
   &+ \e^{-\int_x^a \rmi\mathcal{A}_x^{(\nu)}(x') dx'} c^{(\nu)}(a) \, \sin k^{(\nu)}x
   \Big] / \sin k^{(\nu)}a\,.
\end{align}
%
%
In the absence of a magnetic field, we introduce the Bloch wave vectors $\bm q = (q_x,q_y)$ so that $ c^{(\nu)}(ma,na)= c^{(\nu)} \e^{\rmi a (m q_x+ n q_y)}$. Then for the triplet modes the dispersion relation assumes the form 
$\bm H_{\bm q} \bm c^{(t)} =  \bm c^{(t)}\,\cos k^{(t)}a$, where 
$\bm H_{\bm q} = (\bm H^{(x)}_{q_x}+ \bm H^{(y)}_{q_y})/2$,
\begin{align}
  \bm H^{x(y)}_q=  \cos [(q +\rmi \hat{\bm\Lambda}_{x(y)})a]\,,
\end{align}
and $\hat{\bm\Lambda}_{\alpha} \bm c^{(t)} = {\bm\Lambda}_{\alpha} \times \bm c^{(t)}$.
The summation over dispersive triplet modes can be represented as a complex contour integral
\begin{align}
&\delta\sigma^{(t,\rm disp)}_x(x) = \frac{e^2}{2 h}\, \sum_{\bm q} \operatorname{Tr}\oint
\frac{a\,dk}{2\pi \rmi}\, \frac{\sin ka}{\cos ka-\bm H_{\bm q}} \\ \nonumber
&\times\frac{\sin^2 kx  + \sin^2 k(a-x) + 2 \sin kx \sin k(a-x) \, \bm  H^{(x)}_{q_x} }{(k^2a^2+ \kappa^2) \sin^2 ka }  
\end{align}
where the integration contour encloses all $k^{(t)}$ such that $\cos k^{(t)}a$ is an eigenvalue of $\bm H_{\bm q}$, and $\sum_{\bm q} \equiv \int d^2q/(2\pi)^2$. Since the sum of all residues must vanish, we move the integration to all remaining poles of the function: $k = \pm \rmi \kappa/a$ and $k = \pi p/a$ with integer $p$, which gives the two contributions $\delta\sigma^{(t,\rm disp)}_x(x)= \delta\sigma^{(t1)}_x(x)+\delta\sigma^{(t2)}_x(x)$,
\begin{align}\label{eq:xdisp}
&\delta\sigma^{(t1)}_x(x) 
= -\frac{e^2}{2 h}\, \sum_{\bm q} \operatorname{Tr} \frac{1}{\cosh \kappa  - \bm H_{\bm q}}\, \times\\ \nonumber
&\frac{\sinh^2 \frac{\kappa x}a  + \sinh^2  \frac{\kappa(a-x)}a + 2 \sinh  \frac{\kappa x}a \sinh \frac{\kappa(a-x)}a \, \bm  H^{(x)}_{q_x} }{\kappa\,\sinh \kappa} , \\ \nonumber
&\delta\sigma^{(t2)}_x(x) = -\frac{e^2}{h}\, \sum_{\bm q}  \sum_{p=1}^{\infty} \operatorname{Tr}\frac{(-1)^p  -    \bm  H^{(x)}_{q_x}}{(-1)^p -  \bm H_{\bm q}}\, \frac{2 \sin^2 \pi p x }{(\pi p)^2+ \kappa^2  }  \,.
\end{align}

Now we turn to the contribution of the flat cooperon modes, which vanish at the network vertices. They read $\bm c(x,0) = \e^{\hat{\bm\Lambda}_x x}\bm c_{x} \sin ( \pi p x/a)$ and $\bm c(0,y) = \e^{\hat{\bm\Lambda}_y y} \bm c_{y} \sin (\pi p y/a)$, where the amplitudes must satisfy the  condition of zero flux divergence 
\begin{align}
    &(1-\e^{\rmi\pi p-\rmi q_x a + \bm\Lambda_x a}) \bm c_x + (1-\e^{\rmi\pi p -\rmi q_y a + \bm\Lambda_y a}) \bm c_y = 0 
\end{align}
and the normalization condition $|\bm c_x|^2 + |\bm c_y|^2 = 2/(wa)$.
Then, the contribution of the flat triplet modes to the conductivity reads
\begin{align}\label{eq:xflat}
\delta\sigma^{(t3)}_x(x) = -\frac{e^2 }{ h }\, \sum_{\bm q} \sum_{p=1}^{\infty} \operatorname{Tr} \frac{(-1)^p- \bm H^{(y)}_{q_y}}{(-1)^p- \bm H_{\bm q}} \frac{2\sin^2 \frac{\pi p x}{a}}{(\pi p)^2+ \kappa^2 }
\end{align}
Note that the sum of the contributions $\delta\sigma^{(t2)}_x(x) +\delta\sigma^{(t3)}_x(x) = 3\,\delta\sigma_0(x)$ where
\begin{align}
   \delta\sigma_0(x) =  -\frac{2e^2}{h} \, \frac{\sinh  \frac{\kappa x}a  \sinh\frac{\kappa(a-x)}a}{\kappa  \sinh\kappa }
\end{align}
is independent of the spin-orbit interaction. 

Adding the contribution of the singlet modes, averaging over the coordinate, and separating the isotropic and anisotropic parts, we finally obtain 
\begin{align}
  \delta\bar\sigma &=  - \frac{e^2}{h}\Bigg[\frac{\kappa \coth \kappa -1}{\kappa^2} + \frac{\sinh\kappa }{2\kappa} \label{eq:si} \\[-1mm] \nonumber
  & \times \sum_{\bm q}\left(  \operatorname{Tr} \frac{1}{\cosh \kappa  - \bm H_{\bm q}} - \frac{1}{\cosh \kappa  -  H_{\bm q}^{(s)}}\right)\Bigg]\,, \\
  \delta\sigma&_a =  -\frac{e^2}{2h}\, \frac{\kappa \coth \kappa - 1}{\kappa^2 } \sum_{\bm q} \operatorname{Tr} \frac{\bm H^{(x)}_{q_x} - \bm H^{(y)}_{q_y}}{\cosh \kappa  - \bm H_{\bm q}} \label{eq:sa} \,,
\end{align}
where $H^{(s)}_{\bm q} = (\cos q_xa+\cos q_ya)/2$ is the Hamiltonian of the dispersive singlet cooperon modes. The first term in Eq.~\eqref{eq:si} is obtained by the coordinate-averaging of $\delta\bar\sigma_0(x)$ and was denoted as $\delta\bar\sigma_0$ in the main text.

Note that even though $\sigma_a$ does not include any additional small parameters as compared with $\delta\bar\sigma$, except in classes (i), (ii), and (iv), where $\delta\bar\sigma$ has a logarithmic divergence, the typical values of $\sigma_a$ are about two orders of magnitude smaller than those of $\delta\bar\sigma$.

\section{Transfer matrix technique for ballistic networks}\label{app:tm}
\label{App:B}

Now we consider a network that is infinite in the $y$ direction $N_y \to \infty$, but finite along $x$. Then we introduce 
$\psi_{m,n} = \sum_{q_y} \psi_{m,q_y} \e^{\rmi n q_y a}$, where the quasi-wave vector $q_y$ is a good quantum number, since for the chosen magnetic field gauge $A$ does not depend on $y$. Using the Griffith boundary conditions, we relate 
\begin{align}
    \left(\begin{array}{c}
\psi_{m,q_y} \\
\psi_{m+1,q_y}    
\end{array}\right) = T_{n,q_y} 
\left(\begin{array}{c}
\psi_{m-1,q_y} \\
\psi_{m,q_y}    
\end{array}\right)
\end{align}
where the transfer matrix reads
\begin{align}
    &T_{n,q_y} = \left(\begin{array}{cc}
0 & 1 \\
-\e^{\rmi \bm\Lambda_x\cdot\bm\sigma} & 2\e^{\frac{\rmi}2 \bm\Lambda_x\cdot\bm\sigma}(2\cos ka - \cos \tilde q_n a )   
\end{array}\right)
\end{align}
with $\tilde q_{n} = q_y - 2\pi bn - \frac{1}2 \bm\Lambda_y\cdot\bm\sigma$. The transmission matrix through the whole network is then given by $T_{q_y} = T_{N_x,q_y}\ldots T_{2,q_y} T_{1,q_y}$. Now, we switch to the basis of running waves $\psi_{m,q_y}^{\rightarrow(\leftarrow)}$ using the linear transform
\begin{align}
 &\left(\begin{array}{c}
\psi_{m,q_y} \\
\psi_{m+1,q_y}    
\end{array}\right) = L
\left(\begin{array}{c}
\psi_{m\rightarrow,q_y} \\
\psi_{m\leftarrow ,q_y}    
\end{array}\right) 
\end{align}
with
\begin{align}
&L=\left(\begin{array}{cc}
1 & 1 \\
\e^{\rmi (\frac12\bm\Lambda_x\cdot\bm\sigma + k)a} &  \e^{\rmi(\frac12\bm\Lambda_x\cdot\bm\sigma - k)a}  
\end{array}\right) \,.
\end{align}
Then, the transmission coefficient from right to left is obtained as
\begin{align}
    s_{q_y} =  [(L^{-1}T_{q_y} L)_{\leftarrow,\leftarrow}]^{-1}\,,
\end{align}
and the ballistic network conductance reads
\begin{align}
    \sigma = \frac{e^2}{h} \sum_{q_y} \operatorname{Tr}(s_{q_y}^\dag s_{q_y} ) \,.
\end{align}


%

\end{document}